
\magnification 1200
\vsize=38pc
\
\vskip2pc

\centerline{\bf ON MACROSCOPIC ENERGY GAP FOR $q$-QUANTUM MECHANICAL SYSTEMS}

\vskip7pc

\centerline{\bf Ctirad Klim\v c\'{\i}k}


\centerline{Nuclear Centre, Charles University,}

\centerline{V Hole\v sovi\v ck\'ach 2, CS-180 00 Prague 8, \v CSFR}

\vskip1pc

\centerline{\it and}

\vskip1pc

\centerline{\bf Eliza Klim\v c\'{\i}k}


\centerline{Dept. of Math.Analysis, Charles University,}

\centerline{Sokolovsk\'a 83,CS-180 00 Prague 8, \v CSFR}

\vskip1pc

\centerline{\it Dedicated to our daughter Nela at the occasion of her
first birthday.}

\vskip1pc


\vskip1pc

{\bf Abstract:}

\vskip1pc

The $q$-deformed harmonic oscillator within the framework of the recently
introduced
Schwenk-Wess $q$-Heisenberg algebra is considered. It is shown, that for
"physical" values $q\sim1$, the gap between the energy levels decreases with
growing energy. Comparing with the other (real) $q$-deformations of the
harmonic oscillator, where the gap instead increases, indicates that the
formation of the macroscopic energy gap in the Schwenk-Wess $q$-Quantum
Mechanics may be avoided.

\vskip1pc

\centerline{ICP classification number: 0365.}


\par\vfill\supereject

The harmonic oscillator is the typical physical system which the various
$q$-deformations are tested on [1-4]. In this note, we shall concentrate
on one particular aspect of the behaviour of the spectra of the $q$-deformed
oscillators. Namely, the gap between the energy levels increases with
growing energy,  in the models considered in [1-4].
 As already mentioned in [4], one eventually ends up
with the macroscopic gap for the high energy levels, unless the undeformed
case $q=1$ is considered.
 Moreover, the gap between the energy
levels grows exponentially with the energy if $q\neq1$.  It is questionable
whether such behaviour could remain "unseen" in our physical world if
$q$ were not  equal to one.

We shall show , that the $q$-harmonic  oscillator within the framework of the
recently proposed Schwenk-Wess $q$-deformation of the Heisenberg algebra
has the gap between the energy level {\it decreasing} with growing energy.
We shall do it within the framework of the perturbation theory, where the
$q$-deformed Hamiltonian differs from the undeformed one by the perturbation.
This approach should give reliable results, since, anyway, the physics requires
that $q\sim1$.

The Schwenk-Wess $q$-deformation of the Heisenberg algebra [5] was motivated
by the differential calculus on quantum planes [6] and it reads

$$\xi\pi - q^{-1}\pi\xi = {\rm i}\eta,\eqno(1)$$

$$\eta\pi=q\pi\eta, \qquad \eta\xi=q^{-1}\xi\eta, \eqno(2a,b)$$

$$\eta\eta^+=\eta^+\eta=q^{-1},\eqno(3)$$

where $\xi$ and $\pi$ are selfadjoint, $q$ real and $q\ge 1$. The explicit
representations of this algebra were also found in [5]. However, those
representations were not suited for direct limiting $q\rightarrow 1$.
We give another representation of this algebra, given by

$$\pi = {\rm p}{[{\rm ixp}]_q\over {\rm ixp}},\quad \xi={\rm x},\quad \eta=
q^{-{\rm ipx}},\eqno(4)$$

\noindent where x and p are the standard quantum mechanical operators
of the position and of the momentum respectively, obeying the commutation
relations of the undeformed Heisenberg algebra, and the symbol $[.]_q$,
as usual, means

$$[A]_q\equiv {q^A-q^{-A}\over q-q^{-1}}, \eqno(5)$$

\noindent Note, that in $x$-representation the deformed operator $\pi$
is proportional to nothing but the well-known $q$-derivative, i.e.

$$\pi\psi(x)=-{\rm i}D_q \psi(x)\equiv -{\rm i}{\psi(qx)-\psi(q^{-1}x)
\over (q-q^{-1})x}.\eqno(6)$$

\noindent Using the commutation relations of the undeformed Heisenberg
algebra, we find that the operator $\pi$ is self-adjoint\footnote{*}{Rigorously
speaking, we did not study the domains of definition of the operators $\pi$
and $\eta$, given by Eq.(4). Such investigation would be beyond the scope
of this paper.}, as it should. We do not say, however, that the representation
(4)
and the representations given by Schwenk and Wess in Ref.[5] are equivalent.
In fact, we have no reason to assert, that an analogue of the von Neumann
theorem should be valid in the $q$-deformed case. In the case of the standard
(i.e. undeformed) quantum systems with the infinite number of degrees of
freedom, however, not only the algebra of observables but also its
representation
have to be chosen, in order to specify the system [7]. Hence, we shall adopt
the same point of view also in the present case and study the algebra (1-3)
with its representation (4).

We pick up the Hamiltonian of the $q$-harmonic oscillator as follows

$$H={1\over 2}(\pi^2+\omega^2\xi^2), \eqno(7)$$

\noindent where $\omega$ is a real parameter. We wish to identify the spectrum
of (7). The standard combinations $(\pi\pm{\rm i}\omega\xi)/\sqrt{2\omega}$
do not play the role of the raising and lowering operators, however. The method
[4] for constructing the raising and lowering operators for the $q$-deformed
Schr\"odinger problems does not seem to work either, in our case. We may,
however, perform
 the perturbation expansion in the parameter of deformation and look
after  the modifications of the spectrum. The eventual result will
certainly give us the relevant information about the behaviour of the energy
gap,
because, anyway, the physical $q$ cannot differ much from $1$. Set

$$q={\rm e}^{\beta},\eqno(8)$$

\noindent and expand

$$\eqalign{{\rm H}&={\rm H}_0+{\rm H}_1\beta^2+{\rm H}_2\beta^4+\dots\equiv
{1\over 2}({\rm p}^2+\omega^2{\rm x}^2)
+{1\over 2}\big[({\rm ipx})^2{\rm p}^2+{\rm p}^2({\rm ixp})^2-2{\rm p}^2
\big]{1\over 3!}\beta^2\cr &+{1\over 2}
\big[{({\rm ipx})^4\over 5!}{\rm p}^2+{\rm p}^2{({\rm ixp})^4\over 5!}+
{({\rm ipx})^2 {\rm p}^2
({\rm ixp})^2\over 3!3!}-{({\rm ipx})^2{\rm p}^2+{\rm p}^2({\rm ixp})^2
\over 3.3!}+{{\rm p}^2\over 3.5}
\big]\beta^4+\dots \cr}.\eqno(9)$$

\noindent The eigenvalues $E_n(\beta^2)$ and the eigenvectors $\vert \psi_n
(\beta^2)>$
will have the following expansions

$$E_n=E_{n,0}+E_{n,1}\beta^2+E_{n,2}\beta^4+\dots, \eqno(10)$$

$$\vert \psi_n>=\vert\psi_{n,0}>+\vert \psi_{n,1}>\beta^2+\vert\psi_{n,2}>
\beta^4+\dots.\eqno(11)$$

\noindent Expanding the equation

$$({\rm H}-E_n)\vert\psi_n>=0,\eqno(12)$$

\noindent gives

$$({\rm H}_0-E_{n,0})\vert \psi_{n,0}>=0,\eqno(13a)$$

$$({\rm H}_1-E_{n,1})\vert \psi_{n,0}>+({\rm H}_0-E_{n,0})\vert \psi_{n,1}>=0,
\eqno(13b)$$

$$({\rm H}_2-E_{n,2})\vert \psi_{n,0}>+({\rm H}_1-E_{n,1})\vert \psi_{n,1}>
+({\rm H}_0-E_{n,0})\vert\psi_{n,2}>=0,\eqno(13c)$$

$$\dots$$

\noindent It is easy to conclude, that

$$E_{n,1}=<\psi_{n,0}\vert {\rm H}_1 \vert \psi_{n,0}>,\eqno(14a)$$

$$E_{n,2}=<\psi_{n,0}\vert {\rm H}_2\vert \psi_{n,0}>+\sum_{l\neq n}
{{\vert<\psi_{l,0}
\vert {\rm H}_1\vert \psi_{n,0}>\vert^2\over E_{n,0}-E_{l,0}}}.\eqno(14b)$$

\noindent Note, that the first term in the right hand side of (14b) is usually
omitted in the standard textbooks on the Quantum Mechanics, since the expansion
like (9) is usually written as the sum of only two terms, i.e., the unperturbed
Hamiltonian and the perturbation. Hence,e.g., the usual conclusion about the
negativeness
of the second correction to the vacuum energy need not be valid in our case.

The most convenient representation for the evaluation of the required matrix
elements
is the energy representation of the unperturbed oscillator, i.e. the
representation
in which the operators ${\rm a,a^+}$, given by

$${\rm a}={{\rm p}-{\rm i}\omega{\rm x}\over \sqrt{2\omega}},\quad {\rm a^+}=
{{\rm p}+{\rm i}\omega{\rm x}\over \sqrt{2\omega}},\eqno(15)$$

\noindent act as follows

$$ {\rm a}\vert \psi_{n,0}>=\sqrt{n}\vert \psi_{n-1,0}>,\quad {\rm a^+}\vert
\psi_{n,0}>=
\sqrt{n+1}\vert \psi_{n+1,0}>, \eqno(16)$$

\noindent where $n=0,1,\dots$. Inserting the expressions

$${\rm p}={\sqrt{\omega}({\rm a}+{\rm a^+})\over \sqrt{2}},\quad {\rm x}=
{({\rm a^+}-{\rm a})\over {\rm i}\sqrt{2\omega}}\eqno(17)$$

\noindent into (9) we may write, e.g.

$$E_{n,1}={\omega\over 2^4.3!}<\psi_{n,0}\vert \Big[({\rm a^+}^2-
{\rm a}^2+1)^2({\rm a}+{\rm a^+})^2+
({\rm a}+{\rm a^+})^2({\rm a^+}^2-{\rm a}^2-1)^2-8({\rm a}
+{\rm a^+})^2\Big]\vert \psi_{n,0}>.\eqno(18)$$

\noindent Using (16), we have

$$E_{n,1}=-{\omega\over 2.3!}\big\{(n+{1\over 2})^3+{17\over 2^2}
(n+{1\over 2})\big\}.\eqno(19)$$

\noindent We observe, that the first order corrections to the energy levels
are negative and increasing with $n$
in the absolute value. Moreover, the gap between the energy levels tend to
diminish
with growing $n$. The calculation
of the second order corrections require the knowledge of the off-diagonal
matrix
elements of ${\rm H}_1$ as well as of the diagonal elements of ${\rm H}_2$.
A short look at Eq.(9) suggests that the calculation be somewhat lenghty.
Nevertheless, the careful (and several times checked) evaluation gives the
following result for the energy levels up to the second order

$$E_n=\omega (n+{1\over 2})\Big\{1-\beta^2{1\over 2.3!}\big[(n+{1\over 2})^2
+{17\over 2^2}\big]
-\beta^4{7\over 2^3.5!}\big[(n+{1\over 2})^4+{5^2.13\over 2.3.7}
(n+{1\over 2})^2-
{1031\over 2^4.3.7}\big]+\dots\Big\}\eqno(20)$$

\noindent With the exception of the vacuum, the second correction is also
always negative and for large $n$ the second correction to the energy levels
gap decreases with growing $n$.

Needless to say, the result is not very suggestive for seeking after the
exact solution. One requires the better understanding of the algebraic
structure of the Schwenk-Wess algebra, which would provide a construction
of the appropriate raising and lowering operators. We feel, however, that the
perturbative result {\it is} relevant and, more generally, that the
phenomenology of the Schwenk-Wess
$q$-quantum mechanics may be of interest.

\par\vfill\supereject

\vskip3pc

\centerline{\bf References}

\vskip1pc

[1] A.J.Macfarlane, J.Phys. A: Math.Gen. {\bf 22} (1989) 4581;

\vskip1pc

[2] L.C.Biedenharn, J.Phys. A: Math.Gen. {\bf 22} (1989) L873;

\vskip1pc

[3] Chang-Pu Sun and Hong-Chen Fu, J.Phys. A: Math. Gen. {\bf 22} (1989) L983;

\vskip1pc

[4] G.Fiore, $SO_q(N)$-symmetric Harmonic Oscillator on the $N$-dim Real
Quantum

{}~~~~~Euclidean Space,
 SISSA preprint, 35/92/EP (Mar 92);

\vskip1pc

[5] J.Schwenk and J.Wess, A $q$-deformed Quantum Mechanical Toy Model,
Munich

{}~~~~~preprint, MPI-Ph/92-8;

\vskip1pc

[6]  Yu.Manin, Notes on Quantum Groups and Quantum De Rham Complexes,
Bonn

{}~~~~~preprint MPI/91-60 (1991);

\vskip1pc

[7]  F.Strocchi, Elements of quantum mechanics of infinite systems;
World Scientific,

{}~~~ Singapore (1985)

\end